\documentclass{article}
\usepackage{spconf,amsmath,graphicx}

\title{Interpreting intermediate convolutional layers in unsupervised acoustic word classification}
\name{Ga\v{s}per Begu\v{s} and Alan Zhou\thanks{This research was funded by faculty development grants at UC Berkeley and the University of Washington. We would like to thank Sameer Arshad for slicing data from TIMIT.}}
\address{University of California, Berkeley%\\\texttt{\{begus,azhou314\}@berkeley.edu}
}
\usepackage{amsfonts}
\usepackage{amssymb}
\usepackage[safe]{tipa}

\begin{document}
\ninept
\maketitle
\begin{abstract}

Understanding how deep convolutional neural networks classify data has been subject to extensive research. This paper proposes a technique to visualize and interpret intermediate layers of unsupervised deep convolutional networks by averaging over individual feature maps in each convolutional layer and inferring underlying distributions of words with non-linear regression techniques. A GAN-based architecture (ciwGAN \cite{begusCiw}) that includes a Generator, a Discriminator, and a classifier was trained on unlabeled sliced lexical items from TIMIT. The training process results in a deep convolutional network that learns to classify words into discrete classes only from the requirement of the Generator to output informative data. This classifier network has no access to the training data -- only to the generated data. We propose a technique to visualize individual convolutional layers in the classifier that yields highly informative time-series data for each convolutional layer and apply it to unobserved test data. Using non-linear regression, we infer underlying distributions for each word which allows us to analyze both absolute values and shapes of individual words at different convolutional layers, as well as perform hypothesis testing on their acoustic properties. The technique also allows us to test individual phone contrasts and how they are represented at each layer.

\end{abstract}
\begin{keywords}
unsupervised acoustic word embedding, interpretable deep learning, ASR, generative models, generalized additive mixed models
\end{keywords}
\section{Introduction}

Several prominent speech processing models involve deep convolutional networks, both in supervised \cite{palaz13,Abdel-HamidCNN,zhang16b_interspeechCNN} and unsupervised  settings \cite{baevski20,niekerk20,chorowski19}.  Models using CNNs achieve high performance for phone and vocabulary recognition tasks, but substantially fewer studies focus on interpretability of the networks.

\subsection{Prior work}

Interpreting and visualization of intermediate convolutional layers in CNNs has primarily focused on the visual domain  \cite{zeiler14}, but techniques have recently been proposed for speech data as well \cite{huang15,krug19,muckenhirn19,chowdhury20}. Most proposals focus on visualizing and interpreting filters of supervised models \cite{ravanelli19,krug18introspection,krug18,krug19,huang15,palaz15,golik15}. \cite{krug18introspection} visualize filters of models trained on spectrograms; \cite{muckenhirn19} focuses on relevance maps using gradient-based visualization. \cite{millet21} focuses on activation maps and compare them to brain data. \cite{belinkov17} cluster feature representations of each layer in a DeepSpeech model. \cite{krug18,krug19} involve feature maps, but use a different technique that averages entire feature maps across separate spectrogram inputs and focus on activation profiles for individual neurons. \cite{harwath19} also use feature maps, but take the L2 norm of convolutional feature maps of a spectrogram, focusing on peak activity and segment boundaries in one convolutional layer rather than on any acoustic property across all layers.

\subsection{Goals}
 Unlike most previous proposals, our work uses unsupervised models trained on raw acoustic data. We are primarily interested in how representation of linguistically meaningful units self-emerges in unsupervised generative models. Additionally, instead of on filters, we focus on visualizing and averaging feature maps and argue that they yield highly informative time-series data that allows acoustic analysis of any acoustic property in the intermediate layers and requires no other clustering techniques (as in \cite{harwath19}).  Finally, we introduce non-linear regression techniques to the study of learned intermediate representations in CNNs.

This work also falls in line of the larger scope of unsupervised acoustic word embedding \cite{kamper14,chung16,shain19,baevski20,niekerk20,chorowski19}. The majority of proposals in this framework operate with variational autoencoders (VAEs). This paper is thus part of a larger attempt to build interpretable unsupervised acoustic word embedding models within the GAN framework \cite{begus19,begusCiw}.

\section{Models \& techniques}

\subsection{Models}

We use Categorical and Featural InfoWaveGAN (ciwGAN and fiwGAN) \cite{begusCiw}. CiwGAN/fiwGAN is an  InfoGAN \cite{chen16} extension  of the WaveGAN \cite{donahue19} architecture which, unlike the InfoGAN proposal, features a separate Q-network that can model lexical learning. Like WaveGAN (based on \cite{radford15} and \cite{goodfellow14}), the model consists of two networks, a Generator  $G$ which attempts to generate audio samples given a latent distribution $z$, and a Discriminator $D$ which takes both Generator outputs $G(z)$ and real data $x$ as inputs and attempts to estimate the Wasserstein distance between the input distribution and the real distribution. Just as in WaveGAN, the Generator uses 1D transpose convolutions to upsample from a low-dimensional latent space to audio, while the Discriminator uses regular 1D convolutions to stride along the audio. FiwGAN differs from ciwGAN in the structure of the latent code: ciwGAN takes a one-hot vector as the latent code ($c$), while  fiwGAN introduces a new latent space structure: binary codes that allow featural learning \cite{begusCiw}.

\begin{figure}
    \centering
    \includegraphics[width=.32\textwidth]{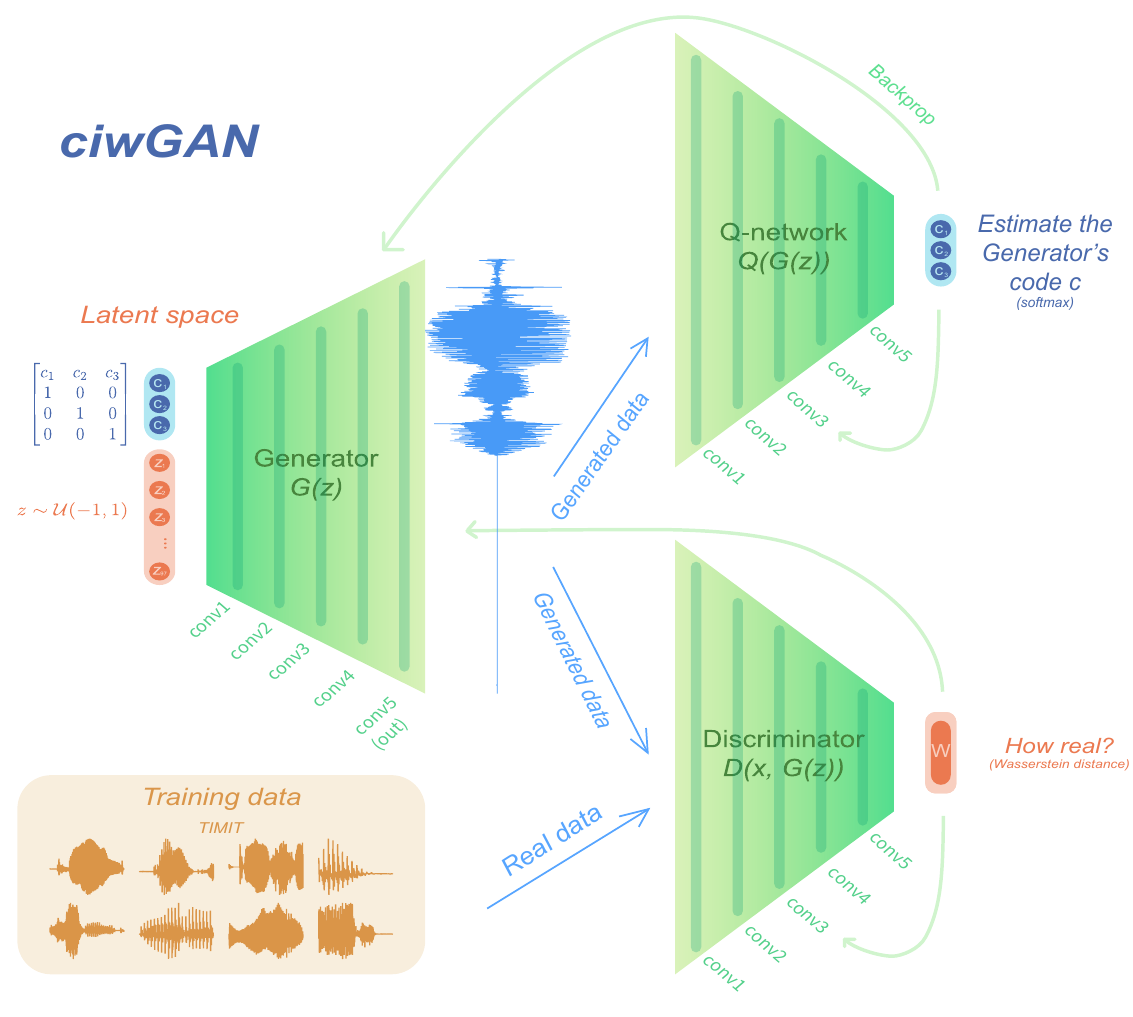}
    \includegraphics[width=.35\textwidth]{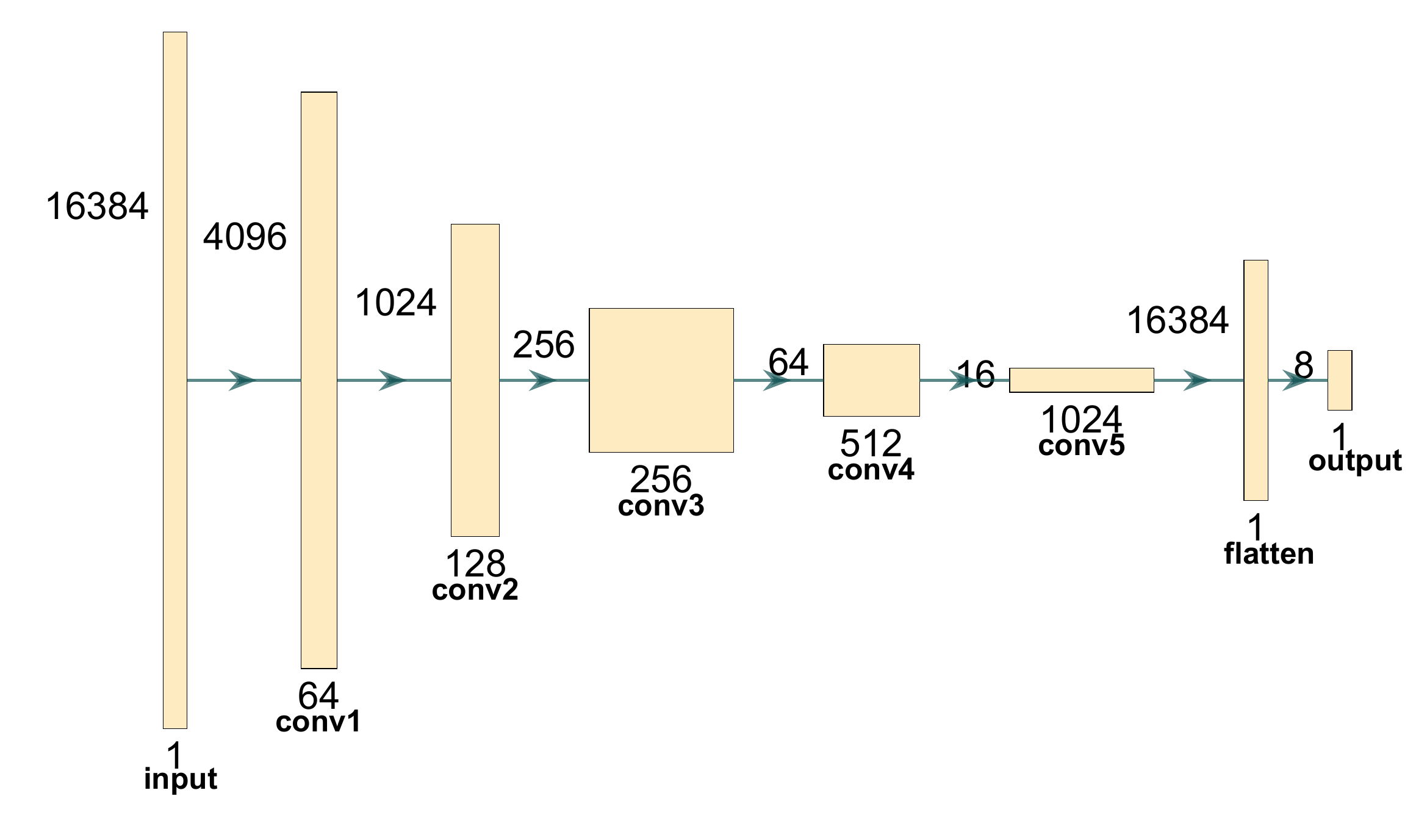}
    \caption{(top) The architecture of the ciwGAN network \cite{begusCiw} (based on \cite{radford15,donahue19,chen16}) schematized for three code variables $c$. A network with eight code variables $c$ was used in training. (bottom) The structure of the Q-network \cite{begusCiw}  with intermediate convolutional layers based on \cite{donahue19} (drawn with \cite{plotneuralnet}).}
    \label{fig:icassp}
\end{figure}

Following \cite{chen16}, we aim to maximize the mutual information between specific latent codes $c$ and the generated outputs $G(z, c)$ by forcing an auxiliary distribution $Q(c|G(z, c))$ to be as close to the true distribution of $P(c|G(z, c))$ as possible. In InfoGAN, this is done by borrowing convolutional layers from the Discriminator to estimate the distribution $Q$. In ciwGAN/fiwGAN, however, we make use of an entirely separate Q-network that shares the convolutional structure (but not weights) of the Discriminator, the only difference being the dimensionality of the final layer. In addition to inducing the use of latent codes in the Generator, the Q-network can be seen as a classification network that aims to categorize audio into latent codes.

Separating the Q-network (on which we perform the visualization technique) from the Discriminator comes with several advantages. First, while the Discriminator comes into contact with both real and generated samples over the course of training, the Q-network only ever comes into contact with generated samples. This means that the Q-network never sees ``real" data, making it unlikely to overfit so long as the Generator continues to produce diverse output. In addition, separating the Q-network follows from the differing objectives of the Q-network and Discriminator. While the Discriminator aims to estimate the input's distance from a distribution, the Q-network aims to categorize the input among several classes. This means that we can perform interpretation and visualization techniques on a classification network trained in an unsupervised manner.

\begin{figure}
    \centering
    \includegraphics[width=0.48\textwidth]{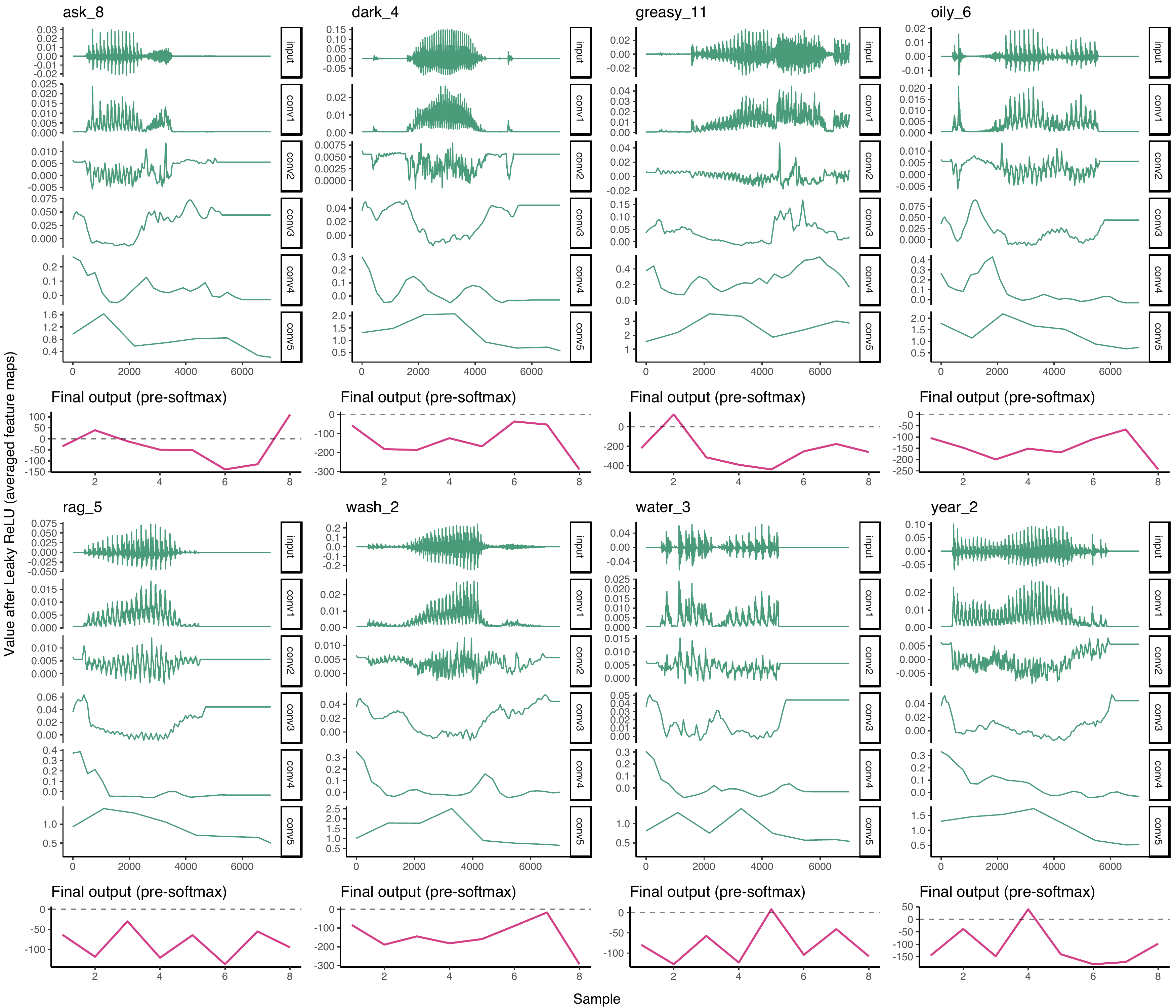}
    \caption{Averaged and upsampled  activations (0-6000) in each layer of the Q-network on an instance of each word from the test set. For each layer, going from top to bottom, we see activations in order from the earliest layers to the latest layers (green), followed by output of the final unnormalized logits in the last layer (purple). While earlier activations follow the stimulus quite closely, later activations become more abstract representations.  }
    \label{fig:drawLayersAll8}
\end{figure}

WaveGAN also \cite{donahue19} introduces  the phase shuffle technique. As the Generator makes exclusive use of transpose convolutions, the outputs that it generates contain periodic artifacts. As these artifacts often occur at the same phase, it is easy for the Discriminator to learn these artifacts, making the Discriminator's training objective trivial. To combat this, WaveGAN makes  shifts the activations in each layer of the Discriminator by a random number of samples. We also make use of phase shuffle in the Q-network for the sake of consistency with the Discriminator. 

\subsection{Visualization techniques}
Following \cite{begusZhou} which focuses exclusively on the Generator, we propose that averaging over individual feature maps in each convolutional layer after the Leaky ReLU activation in the Q-network yields highly interpretable time-series data that summarizes which linguistic properties are encoded at which layer. Specifically, for a convolutional layer $C$, we are able to obtain time-series data $t$ via
\begin{equation}
    t = \frac{1}{\|C\|} \sum_{i=1}^{\|C\|} C_i\label{eq1}
\end{equation}

where $\|C\|$ is the total number of feature maps (or equivalently, the number of channels in the convolutional layer) and $C_i$ is the $i$th feature map.

We apply this technique on the Q-network by feeding the network new data withheld from training. We took samples of the word-slices withheld from the training process, and normalized them to fit with the scale of the Generator output. These normalized samples were then passed through the Q-network with phase shuffle disabled. The activations of each layer are averaged as per (\ref{eq1}). 

To infer the underlying distribution for each word in a given convolutional layer, we fit raw visualizations (averaged feature maps after Leaky ReLU) to non-linear regression models: generalized additive mixed models (GAMMs; \cite{mgcv,soskuthy17}). This allows us to analyze both absolute values (parametric terms) and shapes (smooths) of acoustic words at various convolutional layers and to perform hypothesis testing on differences between words at different layers. 

The proposed technique thus allows testing with inferential statistics of which linguistic contrast are encoded at what layer in an unsupervised deep convolutional network in which learning of linguistically meaningful representations needs to emerge in an unsupervised manner. 

The technique also allows testing of how individual sounds or contrasts (such as difference between the two fricatives [s] and [\textipa{S}] in \textit{ask} and \textit{wash}) are encoded. To test individual contrasts, we annotate phone boundaries in the input and visualize intermediate layers  only for the interval that corresponds to the presence of that feature in the input (Section \ref{var}).

\section{Experiments}

\subsection{CiwGAN}

The first network (ciwGAN) is trained on eight unlabeled single words from TIMIT \cite{timit}: \textit{ask}, \textit{dark}, \textit{greasy}, \textit{oily}, \textit{rag}, \textit{wash}, \textit{water}, and \textit{year}. The smaller number of frequently attested non-function training words is chosen to increase interpretability. 

Four-fifths (80\%) of the sliced TIMIT words  (altogether 4,052 tokens) were used in training; the remaining 20\% (altogether 1,067 tokens) were withheld from the training process as test data.  The words were sliced into individual files and left padded with 25ms window of silence. In order to have inputs that match the fixed length input dimensions of the Discriminator, they are additionally right padded with silence during training so that each item contains 16,384 data points (approximately 1s with 16kHz sampling rate). As the amount of right pad ranges from 4,176 to 13,203 samples, the network still needs to form coherent representations for the entire length of the input.

The model is trained with 8 categorical codes $c$ --- one for each word --- for 121,116 steps, after which collapse (common in GAN training) was observed.

\subsection{Visualization \& regression}
\label{var}

To visualize how representations of words self-emerge in intermediate convolutional layers of the Q-network, we feed 1,067 test items (withheld entirely from the training) to the Q-network in the ciwGAN architecture and average over feature maps at each convolutional layer. 

For the purposes of visualization against the original stimulus, we also upsample outputs at each layer to 16,834 samples using linear interpolation. Examples of these averaged activations for individual words are shown in Figure \ref{fig:drawLayersAll8}.

Raw visualizations reveal that acoustic properties in the input are encoded relatively locally in the first few convolutional layers. For example, the visualization of \textit{dark} in Figure \ref{fig:drawLayersAll8} shows how bursts are encoded with positive values in Conv5, but with a depression in Conv4.

To infer underlying shapes of each lexical item, we fit raw visualizations into a generalized additive mixed model  with Value at Conv5 as the dependent variable; Word identity (treatment-coded with \textit{ask} as reference) as the parametric term; and thin plate smooths for the Value of sample with by-word difference smooth, by-token random smooths, and correction for autocorrelation. We perform the analysis for the fifth convolutional layer because of its highly reduced dimensionality.

\begin{figure*}
    \centering
    \includegraphics[width=.36\textwidth]{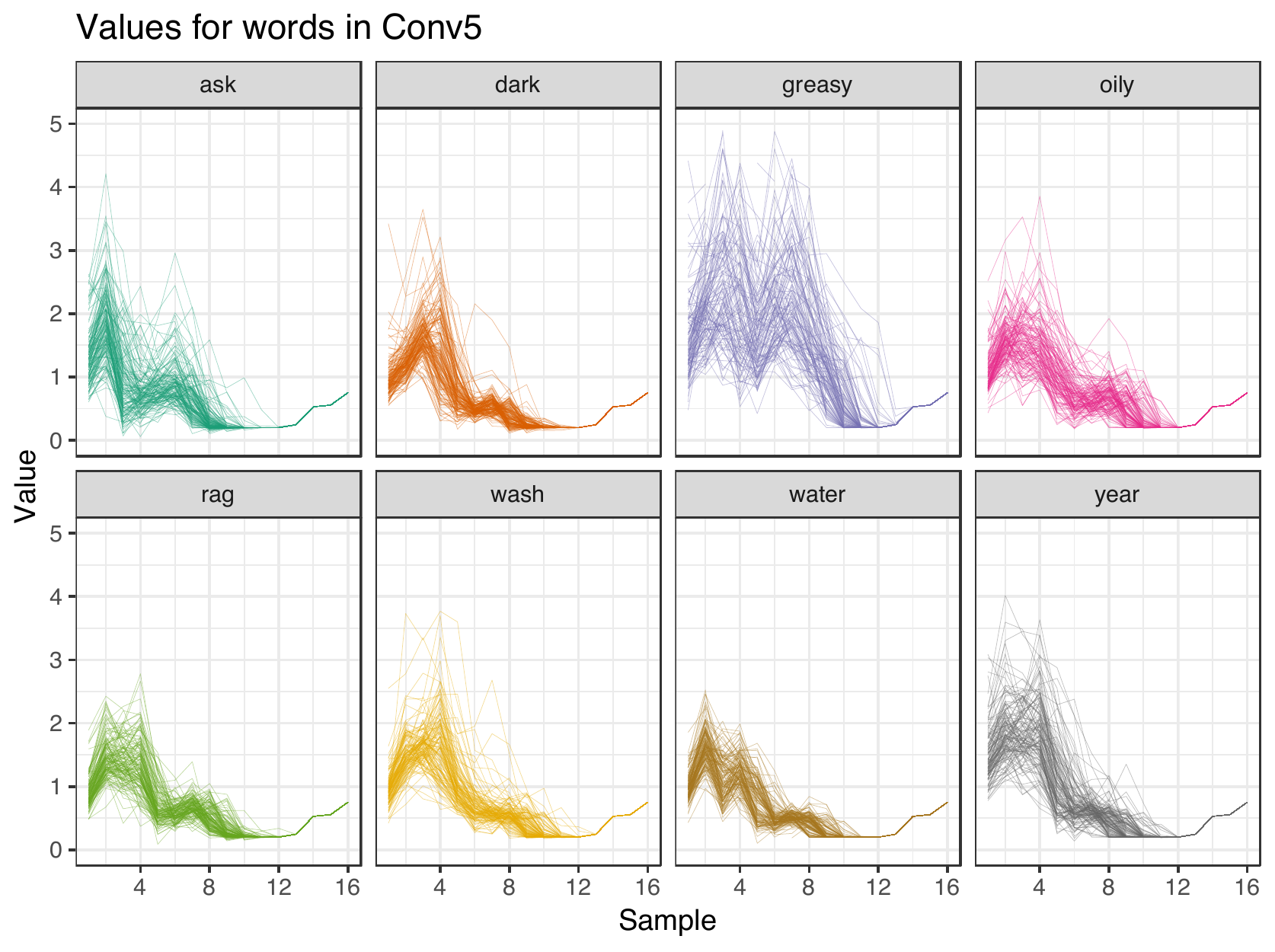}\includegraphics[width=.36\textwidth]{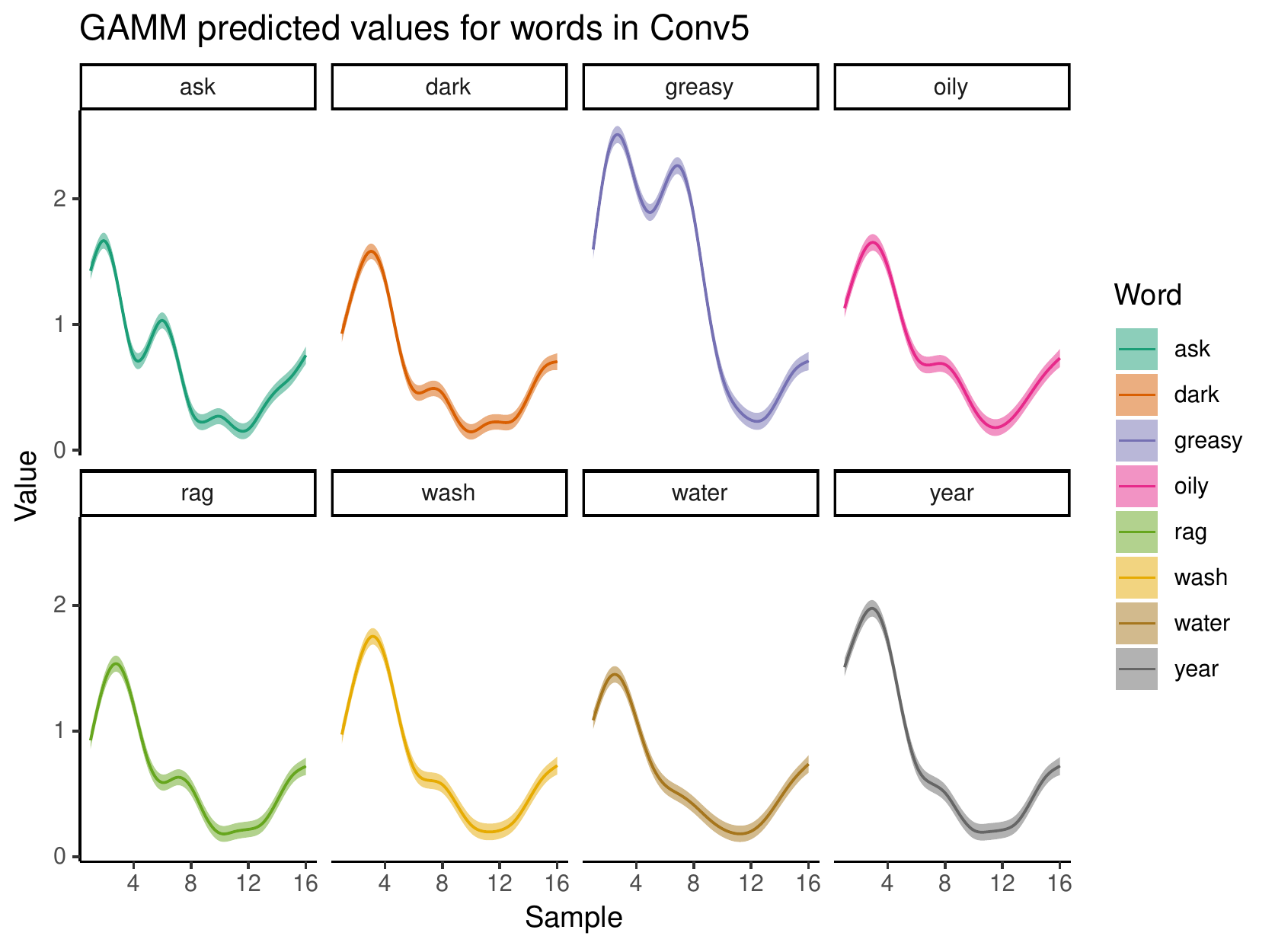} \includegraphics[width=.26\textwidth]{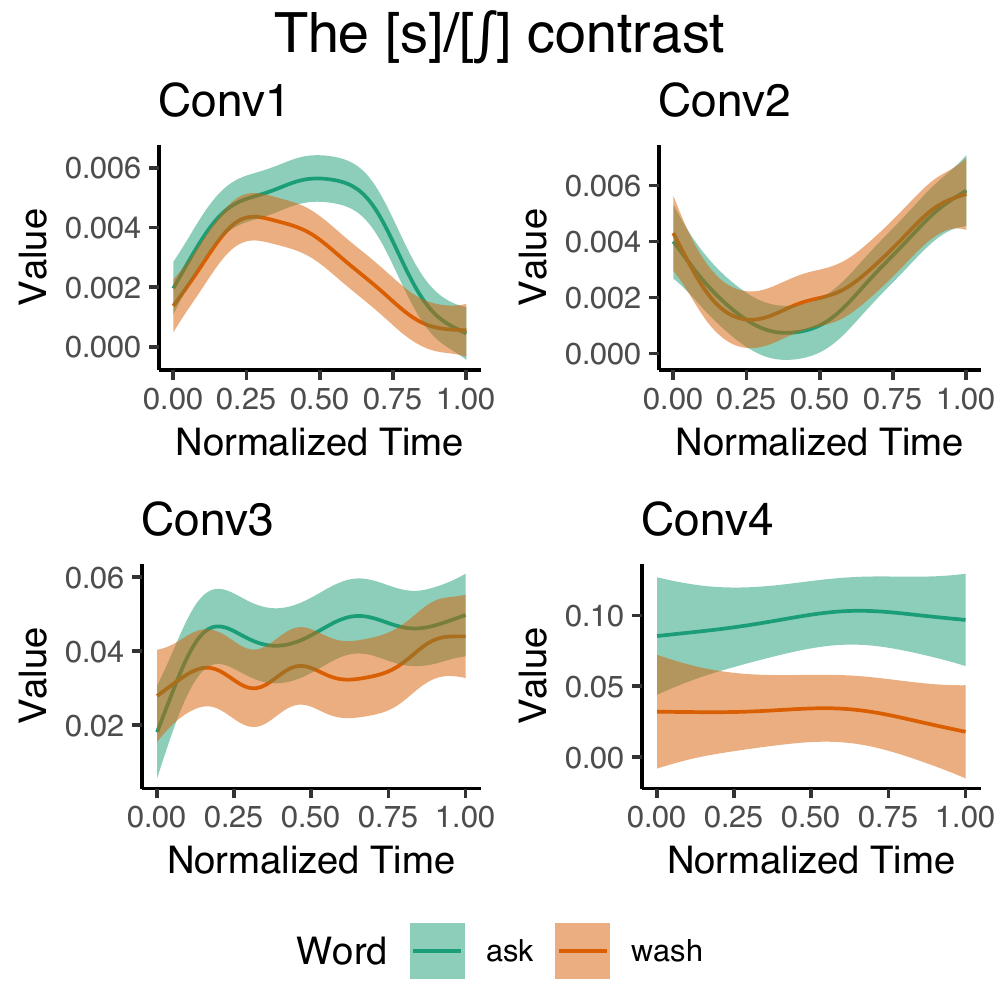}
    \caption{(left) Raw values of the fifth convolutional layer (Conv5) for all tested words across the eight unique words (three datapoints are above the plot's upper limit). (middle) Predicted values of a Generalized Additive Mixed Model (GAMM) \cite{mgcv,soskuthy17}. (right) Predicted values from a GAMM model at the first four convolutional layers for slices that correspond to the fricative part [s]/[\textipa{S}] of \textit{ask} and \textit{wash} in the input.}
    \label{fig:my_label}
\end{figure*}

% latex table generated in R 3.6.3 by xtable 1.8-4 package
% Tue Sep 28 01:44:30 2021
\begin{table}[ht]
\centering
\scalebox{0.75}{\begin{tabular}{lrrrr}
   \hline \hline
A. parametric coefficients & Estimate & Std. Error & t-value & p-value \\  \hline
  (Intercept) & 0.6832 & 0.0215 & 31.7138 & $<$ 0.0001 \\ 
  Word=dark & -0.0409 & 0.0300 & -1.3626 & 0.1730 \\ 
  Word=greasy & 0.6323 & 0.0314 & 20.1116 & $<$ 0.0001 \\ 
  Word=oily & 0.0811 & 0.0312 & 2.6021 & 0.0093 \\ 
  Word=rag & -0.0240 & 0.0305 & -0.7870 & 0.4313 \\ 
  Word=wash & 0.0603 & 0.0311 & 1.9425 & 0.0521 \\ 
  Word=water & -0.0464 & 0.0307 & -1.5126 & 0.1304 \\ 
  Word=year & 0.1298 & 0.0311 & 4.1712 & $<$ 0.0001 \\ 
   \hline
B. smooth terms & edf & Ref.df & F-value & p-value \\  \hline
  s(Sample) & 8.9700 & 8.9934 & 235.5282 & $<$ 0.0001 \\ 
  s(Sample):Word=dark & 8.9684 & 8.9936 & 159.7072 & $<$ 0.0001 \\ 
  s(Sample):Word=greasy & 8.9446 & 8.9886 & 231.4914 & $<$ 0.0001 \\ 
  s(Sample):Word=oily & 8.9562 & 8.9911 & 122.4444 & $<$ 0.0001 \\ 
  s(Sample):Word=rag & 8.9534 & 8.9904 & 110.0716 & $<$ 0.0001 \\ 
  s(Sample):Word=wash & 8.9625 & 8.9924 & 165.8256 & $<$ 0.0001 \\ 
  s(Sample):Word=water & 8.9309 & 8.9855 & 71.5510 & $<$ 0.0001 \\ 
  s(Sample):Word=year & 8.9613 & 8.9922 & 140.2448 & $<$ 0.0001 \\ 
  s(Sample,UniqueWord) & 5327.3250 & 9595.0000 & 4.8275 & $<$ 0.0001 \\ 
   \hline \hline
\end{tabular}}
\caption{Regression estimates of the model in Figure \ref{fig:my_label} (middle).} 
\label{tab.gam}
\end{table}

Estimates of the non-linear regression allow hypothesis testing on both the absolute values and shapes of words at different layers. The parametric coefficients in Table \ref{tab.gam} estimate the absolute values of individual lexical items and their differences: value of \textit{ask} is significantly different from 0 and differs significantly in absolute values  from \textit{greasy}, \textit{oily}, and \textit{year}, but not from other four words (see Table \ref{tab.gam}). In shape,  however, which is estimated with thin plate smooths, \textit{ask} differs significantly from all seven other words. Visualization of the thin plate smooths for each word suggest that the network represents each lexical item with a distinct shape in Conv5 which is then transformed into the final layer (Figure \ref{fig:drawLayersAll8}) that, after softmax, classifies each word into one of the eight classes.

To test how individual sounds are encoded across the convolutional layers, we manually annotated the period  of frication noise of [s] and [\textesh] in 120 instances of two lexical items, \textit{ask} and \textit{wash} (60 each). We extract those values of averaged feature maps (as per \eqref{eq1}) that correspond to the period of frication noise in the input and normalize time. We analyze layers up to the fourth layer (Conv1-4) as Conv5 features too few variables. We fit the data to a generalized additive mixed model.

%\begin{figure}
   % \centering

  %  \caption{}
 %   \label{fig:askwashSGamARConv1ggplotPlotAll}
%\end{figure}

The visualization of the [s]/[\textesh] contrast in Figure \ref{fig:my_label} suggest that the underlying distribution is encoded with a substantial difference in shapes of the two sounds in the first layer. In the second layer (Conv2), the difference in shapes ceases to be significant. Conv3 and Conv4 illustrate how a shape distinction translates into an  absolute value distinction: at Conv4, the shapes of the two sounds are not significantly different ($F=0.35,p=   0.56$), while their absolute values differ significantly ($\beta=-0.068,t=-4.83,p<0.00001$).

\subsection{FiwGAN on entire TIMIT}
To test how the proposed technique scales to larger corpora, we analyze intermediate layers in a fiwGAN network trained on all words from TIMIT (a pretrained network from \cite{begusCiw}):  54,378 tokens of 6,229 unique words. The fiwGAN model allows for a highly reduced vector representation of lexical items \cite{begusCiw}. The fiwGAN network is trained with 13 latent feature variables ($\phi$) which enables $2^{13} = 8,192$ unique classes. Lexical learning emerges despite the mismatch between the unique lexical items in TIMIT (6,229) and the number of unique classes allowed by the architecture (8,192) (for error rates of the Generator, see \cite{begusCiw}).   Unlike in the 8-word ciwGAN model, the test data in the fiwGAN model is not withheld from training. However, as the Q-network only ever accesses Generator outputs during training, we are still testing the Q-network on unseen data, and any potential overfitting should have minimal influence over the results.

The same test data as used for the 8-word ciwGAN model is used in the fiwGAN model for three words: \textit{ask}, \textit{greasy}, and \textit{wash}. 391 tokens of the three words are fed to the model. Averaged feature maps after Leaky ReLU from Conv5 are fit to a generalized additive model. Absolute values of \textit{ask} are significantly different from zero ($\beta=-0.095,    t=  -9.16, p<0.00001$) and from absolute values of \textit{greasy} ($\beta=0.034, t= 2.25, p=  0.024$), but not from \textit{wash} ($\beta=-0.013,   t= -0.88, p=  0.38$). The shapes of both \textit{wash} and \textit{greasy}, however, differ significantly from \textit{ask} as estimated with smooth terms: $ F=  90.3, p<0.00001$ for \textit{greasy} and $ F=  22.9, p<0.00001$ for \textit{wash}.

\begin{figure}
   \centering \includegraphics[width=.34\textwidth]{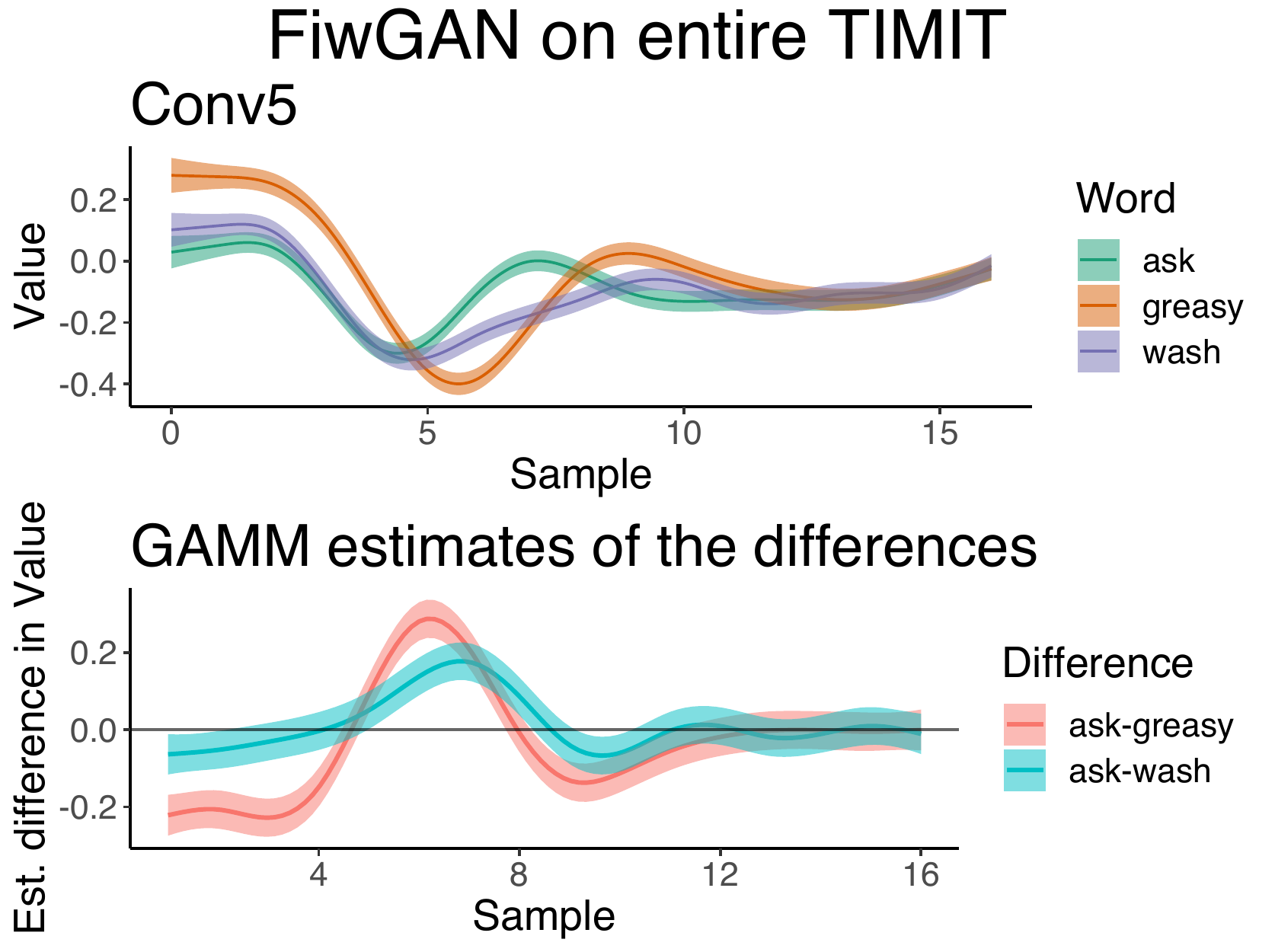}
    \caption{(top) Predicted Conv5 values of the fiwGAN model trained on the entire TIMIT based on a GAMM model for each of the three words tested. (bottom) Difference plot estimating differences between pairs of words.}
    \label{fig:uniqueFiwFilesDfGamARBamGamDiffwashgreasyPlotDiffCOMBINED}
\end{figure}

The difference plot in Figure \ref{fig:uniqueFiwFilesDfGamARBamGamDiffwashgreasyPlotDiffCOMBINED} estimates at which sample points a pair of words differs significantly. We observe that the shapes for \textit{ask} and \textit{wash} differ substantially only in a relatively narrow band that likely corresponds to the identity of the fricative, because that is the most salient acoustic difference between the two words. Like in the ciwGAN model, the values for [\textipa{S}] are significantly lower than values for [s].

\section{Discussion \& conclusion}

Interpreting and visualizing how deep neural networks classify data has primarily focused on the visual domain. Here, we propose a technique to interpret and visualize intermediate convolutional layers when networks learn to classify words from unlabeled data in an unsupervised manner without ever having access to the actual training data in the ciwGAN/fiwGAN framework. This means learning representations of linguistically meaningful properties (such as words) needs to self-emerge in our models. We focus on applying inferential statistics --- generalized additive mixed models --- to infer underlying distributions of word representations. This allows inferential statistical tests of both absolute values and shapes of word representations at each convolutional layer.

The proposed technique opens up several ways to explore and interpret the relationship between the input, the latent space, and intermediate convolutional layers in unsupervised acoustic word embedding tasks. Any acoustic contrast can be tested and compared, both using smaller controlled  settings that are more interpretable (e.g.~models trained on a subset of words) as well as using models trained on entire speech corpora. The technique has the potential to serve as a diagnostic for detecting layers at which speech contrasts (such as phonemes) fail to get encoded. There are several further directions this work should take: from performing acoustic analysis on spectra of the averaged feature maps to exploring encoding of further phonemic contrasts in speech.

%\clearpage

\bibliographystyle{IEEEbib}
\bibliography{bibliography.bib}

\end{document}